\documentclass{article}
\usepackage{graphicx, amsmath, amsthm, amssymb, setspace}
\usepackage[round, sort&compress]{natbib}
\newcommand{\appsection}[1]{\let\oldthesection\thesection\renewcommand{\thesection}{Appendix \oldthesection}\section{#1}\let\thesection\oldthesection}

\begin{document}

\begin{titlepage}

\begin{center}

{\LARGE \bf The Randomized CRM: An Approach to Overcoming the Long-Memory Property of the CRM} \\

\ \\

{\large Joseph S. Koopmeiners$^{1, \dagger}$ and Andrew Wey$^{1}$}\\
\ \\
$^{1}$Division of Biostatistics, University of Minnesota, Minneapolis, Minnesota \\
\ \\
{\large \it $^{\dagger}$Corresponding author's address: koopm007@umn.edu} \\

{\large Technical Report: \today}

\end{center}

\newpage

\begin{abstract}
\noindent The primary object of a phase I clinical trial is to determine the maximum tolerated dose (MTD). Typically, the MTD is identified using a dose-escalation study, where initial subjects are treated at the lowest dose level and subsequent subjects are treated at progressively higher dose levels until the MTD is identified. The continual reassessment method (CRM) is a popular model-based dose-escalation design, which utilizes a formal model for the relationship between dose and toxicity to guide dose-finding. Recently, it was shown that the CRM has a tendency to get ``stuck'' on a dose-level, with little escalation or de-escalation in the late stages of the trial, due to the long-memory property of the CRM. We propose the randomized CRM (rCRM), which introduces random escalation and de-escalation into the standard CRM dose-finding algorithm, as an approach to overcoming the long-memory property of the CRM. We discuss two approaches to random escalation and de-escalation and compare the operating characteristics of the rCRM to the standard CRM by simulation. Our simulation results show that the rCRM identifies the true MTD at a similar rate and results in a similar number of DLTs compared to the standard CRM, while reducing the trial-to-trial variability in the number of cohorts treated at the true MTD.
\end{abstract}

\end{titlepage}

\newpage
\setcounter{page}{3}

\section{Introduction}

The primary objective of phase I clinical trials is to evaluate the safety of a novel therapeutic agent and identify the maximum tolerated dose (MTD). The MTD is defined as the maximum dose with probability of dose limiting toxicity (DLT) less than some pre-specified threshold (typically 0.33 in phase I oncology trials). The MTD is identified in a phase I clinical trial using a dose escalation study, where initial subjects are treated at the lowest dose level and subsequent subjects are treated at progressively higher dose levels until the MTD is identified. A wide variety of dose escalation designs for identifying the MTD have been discussed in the statistical literature.

Phase I dose escalation designs can be broadly classified as either rule-based or model-based designs. Rule-based designs, of which the 3+3 design is the most common example \citep{Storer89}, utilize a simple algorithm for dose escalation where subjects are assigned a dose based on the outcomes for the previous subject or cohort. Other examples of rule-based designs include the broad class of up-and-down designs \citep{SandF02, IMMandD03, OandH09}. Model-based designs, such as the continual reassessment method (CRM) \citep{OPandF90}, require the specification of a (typically parametric) model for the relationship between dose and the probability of toxicity and dose escalation is guided by current estimates of the probability of toxicity at each dose. In the CRM, subjects are treated at the current estimate of the MTD under some restrictions described by \citet{GZandP95}. Other examples of model-based designs include escalation with overdose control (EWOC) \citep{BRandZ98} and the EffTox design \citep{TandC04}, which considers the trade-off between toxicity and efficacy in phase I.

\citet{OandH13} recently investigated the long-memory property of several phase I clinical trial designs, including the CRM. They illustrated that the CRM has a tendency to get ``stuck'' on a dose-level after only a few cohorts and that there is little escalation or de-escalation late in the trial. Due to this phenomenon, the CRM has substantial trial-to-trial variability in the number of subjects treated at the MTD and any single trial may only treat a small number of subjects at the MTD even though the CRM, on average, treats a higher number of subjects at the MTD than other designs. This places practicioners in the difficult position of having to choose between a design that treats a high number of subjects at the MTD, on average, but has high trial-to-trial variability and a design that treats a lower number of subjects at the MTD, on average, but has smaller trial-to-trial variability.

In this manuscript, we propose the randomized CRM (rCRM) and discuss how it can be used to overcome the long-memory property of the CRM. In the CRM, initial cohorts are treated at the lowest dose level and subsequent cohorts are treated at the current estimate of the MTD under some restrictions when escalating. For the rCRM, we alter the standard CRM dose-finding algorithm to allow random escalation or de-escalation in the event that the current estimate of the MTD is the same as the estimate of the MTD before the previous cohort was enrolled. We investigate two schemes to random escalation/de-escalation where the probability of escalation/de-escalation is proportional to the posterior probability that each dose is the MTD. This is a natural approach to overcoming the long-memory property of the CRM and has the added benefit of acknowledging the uncertainty in our estimate of the MTD at any point in the trial. Our simulation results illustrate that the rCRM decreases the trial-to-trial variability in the CRM without dramatically decreasing the probability of correctly identifying the MTD and average number of subjects treated at the MTD.

The remainder of this manuscript proceeds as follows. In Section~\ref{CRM_sec}, we provide a brief overview of the CRM and introduce the rCRM as an approach to overcoming the long-memory property of the CRM. In Section~\ref{sim_sec}, we present simulation results evaluating the operating characteristics of the rCRM and illustrate how the rCRM decreases the trial-to-trial variability in the number of subjects treated at the MTD compared to the standard CRM. Finally, we conclude with a brief discussion in Section~\ref{disc_sec}.

\section{Study Design}
\label{CRM_sec}

In this section, we provide a brief overview of the CRM and then go on to introduce the rCRM. Let $X$ be the primary outcome, DLT, which is a Bernoulli random variable with probability $\theta\left(d\right)$ at dose level $d$. The CRM is a model-based method and requires that a model be specified for the relationship between dose and the probability of DLT. A simple, one-parameter model that is commonly used with the CRM is the power model:
\begin{equation*}
P\left(DLT \vert dose = d\right) = p_{d}^{e^{\alpha}},
\end{equation*}
where $p_{1}, \ldots, p_{D}$ take values between 0 and 1 and are monotonically increasing from $p_{1}$ to $p_{D}$. $\left(p_{1}, \ldots, p_{D}\right)$ is known as the skeleton and must be specified before the study begins. The dose-response relationship for the probability of DLT is controlled by the parameter $\alpha$ and the probability of DLT for all dose levels decreases as $\alpha$ increases. Other, more flexible models can also be used for the association between dose and the probability of DLT, such as a two-parameter logistic model or curve-free methods for modeling the association between dose and the probability of DLT \citep{GandE00}. For the purposes of this manuscript, we will proceed assuming the power model for the relationship between dose level at the probability of DLT but note that changing the model for the probability of DLT would not alter the proposed dose-finding algorithm.

Let $\theta^{*}$ be the pre-specified target probability of DLT that is used to identify the MTD. Typical values for $\theta^{*}$ are 0.20 or 0.33. In the context of the trial, the current estimate of the MTD is the dose level with estimated probability of DLT closest to the target probability of DLT. The CRM begins by treating the first cohort of (typically three) subjects at the lowest dose level. Outcomes for the first cohort are used to update the posterior for $\alpha$ and estimate the MTD. The trial terminates if the posterior probability that the probability of DLT at the lowest dose level exceeds $\theta^{*}$ is greater than some pre-specified threshold, $\pi$. That is, the trial terminates if,
\begin{equation}
\label{safety}
P\left(\theta\left(1\right) > \theta^{*} \vert \vec{X}\right) > \pi.
\end{equation}
Typical values for $\pi$ are 0.90 or 0.95 and $\pi$ is chosen to achieve the desired operating characteristics for the trial. Otherwise, the next cohort of subjects are treated at the dose with estimated probability of DLT closest to the target under the restriction that no dose levels may be skipped when escalating. This continues until the maximum sample size has been reached. The current estimate of the MTD at study completion is declared the MTD. 

\subsection{The Randomized CRM}

A limitation to the standard CRM dose-finding procedure is that it has a tendency to get ``stuck'' on a dose-level early in the trial with very little escalation or de-escalation in the later stages of the trial \citep{OandH13}. As a result, the standard CRM exhibits substantial variability in the number of subjects treated at the true MTD with a large number of subjects treated at the true MTD in some trials but only a small number of subjects treated at the true MTD in others. We will overcome this problem by introducing random escalation or de-escalation to situations where the standard CRM has a tendency to get ``stuck'' on a dose-level.

The rCRM dose-finding algorithm starts in the same manner as the standard CRM dose-finding algorithm. The first cohort of subjects are treated at the lowest dose level and outcomes for the first cohort are used to estimate the posterior for $\alpha$ and estimate the MTD. The trial terminates if there is strong evidence that the lowest dose level is excessively toxic. Otherwise, the next cohort is enrolled at the current estimate of the MTD under the restriction that dose levels may not be skipped when escalating. The difference between the rCRM and the CRM is that the dose will be randomly escalated or de-escalated if the new cohort is to be treated at the same dose-level as the previous cohort. We discuss two approaches to random escalation or de-escalation below. This process continues until the maximum sample size has been reached and the current estimate of the MTD at study completion is declared the MTD. 

We now discuss two approaches to random escalation or de-escalation when the new cohort is to be treated at the same dose-level as the previous cohort. Let $d^{j}$ be the dose level assigned to the $jth$ cohort, $d^{max}$ be the maximum dose level tried in the first $j$ cohorts and $d^{*}$ be the estimate of the MTD after the outcomes are observed for the first $j$ cohorts. In the first randomization scheme (rCRM 1), the next cohort is randomly assigned to $d^{j} - 1$, $d^{j}$ or $d^{j} + 1$ with the following probabilities if $d^{*} = d^{j}$:
\begin{equation*}
P\left(d^{j + 1} = d^{j} + i \right) = \frac { P\left( MTD = d^{j} + i \vert \vec{x} \right) } { \sum_{k = -1}^{1} P\left(MTD = d^{j} + k \vert \vec{x} \right) }
\end{equation*}
for $i = -1, 0, 1$, where $\vec{x}$ are the outcomes for subjects in the first $j$ cohorts. Here, $P\left( MTD = d^{j} + i \vert \vec{x} \right)$ is the posterior probability that dose level $d^{j} + i$ is the true MTD conditional on all available data and we are escalating, de-escalating or staying at the current dose-level with probability proportional to the posterior probability that the three dose-levels are truly the MTD. If $d^{j}$ is the lowest or highest dose level, then $P\left(MTD = d^{j} - 1 \vert \vec{x}\right)$ or $P\left(MTD = d^{j} + 1 \vert \vec{x} \right)$, respectively, are not defined. These probabilities can be treated as being equal to 0 when determining the probability of escalating or de-escalating and the rCRM1 proceeds with no other changes.

The rCRM1 only allows for escalating or de-escalating one dose-level at a time. An alternate randomization scheme (rCRM 2), would be to randomize among all possible dose-levels under the restriction that untried dose-levels can not be skipped. That is, if $d^{*} = d^{j}$, we randomize the next cohort among dose-levels $\{1, \ldots, d^{max} + 1\}$, with the following probabilities:
\begin{equation*}
P\left(d^{j + 1} = i \right) = \frac { P\left( MTD = i \vert \vec{x} \right) } { \sum_{k = 1}^{d^{max} + 1} P\left(MTD = k \vert \vec{x} \right) }
\end{equation*}
for $i = 1, \ldots, d^{max} + 1$. In this case, if $d^{max}$ is equal to the maximum dose level, $D$, we are randomizing subjects to all dose levels with randomization probabilities proportional to the posterior probability that each dose is the MTD.

There are two primary advantages to this approach. First, we eliminate the possibility that the dose-finding algorithm gets ``stuck'' on a dose level by introducing random escalation and de-escalation. Regardless of how many consecutive cohorts have been treated at a single dose level, there will always be some probability of escalating or de-escalating, which makes it unlikely that, say, ten consecutive cohorts would all be treated at the same dose level. Second, the randomized CRM dose-finding algorithm more accurately reflects the uncertainty in our estimate of the MTD. In the standard CRM, subjects are assigned to the current estimate, which is based on only a single summary of the posterior distribution (mean, median, etc.). In contrast, the randomized CRM accounts for the entire posterior distribution and differentiates between precise estimates of the MTD, where there will be little chance of random escalation or de-escalation, and highly variable estimates of the MTD, where there will be a much higher chance of random escalation or de-escalation. As the trial continues and more data are collected, our estimate of the MTD should become more precise and the probability of random escalation or de-escalation should decrease.

In conclusion, our proposed dose finding algorithm is as follows:

\begin{enumerate}
\item Treat the first cohort of m patients at the lowest dose level.
\item Update the posterior distribution for $\alpha$ after outcomes are observed for the first cohort
\item The trial terminates if the lowest dose level has unacceptable toxicity as determined by Equation~\ref{safety}. Otherwise, the next cohort will be treated at the current estimate of the MTD.
\item If this does not result in escalation or de-escalation, the next cohort will be randomly assigned a dose using either the rCRM1 or rCRM2 randomization scheme.
\item The trial continues until termination or until the maximum sample size is reached. The estimated MTD at study completion is declared the MTD.
\end{enumerate}

\section{Simulation Study}
\label{sim_sec}
We completed a small simulation study to evaluate the operating characteristics of the rCRM. Six scenarios were considered, each of which included six dose-levels. For each scenario, we simulated 1000 trials using the CRM, rCRM1 and rCRM2. Each trial had a maximum sample size of 45 subjects with cohorts of three subjects. The target probability of DLT, $\theta^{*}$, was set equal to 0.30 and trials were terminated for overtoxicity if the posterior probability that $\theta\left(1\right) > \theta^{*}$ exceeded $\pi = 0.90$. The association between dose and the probability of DLT was modeled using the power model with a skeleton of $\left(0.01, 0.05, 0.10, 0.18, 0.30, 0.50\right)$. A normal prior with a mean of 0 and standard deviation of 2 was used for $\alpha$. Simulations were completed in R \citep{Rref} and the posterior for $\alpha$ was calculated using JAGS called from R using rjags \citep{rjags}. 

The operating characteristics for the three designs were evaluated by considering the probability of correctly identifying the MTD, the average number of DLTs and the number of cohorts treated at the true MTD. The number of cohorts treated at the true MTD was summarized by both the mean and standard deviation. We consider both the mean and standard deviation, rather than only considering the mean, in order to summarize the trail-to-trial variability of the number of cohorts treated at the true MTD in addition to a measure of central tendency.

We first consider the probability of correctly identifying the MTD and the average number of DLTs for each design. The rCRM must correctly identify the MTD at a similar rate and have a similar safety profile when compared to the standard CRM to be a viable alternative to the CRM. Table~\ref{dose_selection_results} presents the selection probability and the average number of DLTs for the three designs for each of the six scenarios. The two rCRM dose-finding algorithms correctly identify the MTD at a similar rate as the standard CRM. The maximum difference in the probability of correctly identifying the MTD is no more than 0.05 across the six scenarios and trials rarely stop for over toxicity. In addition, there is little difference in the average number of DLTs with less than a one-half DLT difference across designs for each scenario.

\begin{table}[t]
\caption{Simulation results evaluating the operating characteristics of the CRM, rCRM 1 and rCRM 2. Included are the true probability of DLT for each of the six scenarios, the selection probability for the three dose-finding algorithms and the average number of DLTs for the three dose-finding algorithms. 1000 simulations were completed for each scenario.}
\label{dose_selection_results}
\begin{center}
\begin{tabular}{cccccccccc}
\\ \hline
& & & & & & & & & Avg \# \\
& & Dose 1 & Dose 2 & Dose 3 & {\bf Dose 4} & Dose 5 & Dose 6 & Overtoxic & DLTs\\ \hline
{\bf Scenario 1} & P(DLT) & 0.02 & 0.05 & 0.14 & {\bf 0.30} & 0.54 & 0.76 & -- & --  \\
& CRM & 0 & 0 & 0.11 & {\bf 0.78} & 0.11 & 0 & 0 & 12.4 \\
& rCRM 1 & 0 & 0 & 0.13 & {\bf 0.79} & 0.08 & 0 & 0 & 12.7 \\
& rCRM 2 & 0 & 0 & 0.13 & {\bf 0.77} & 0.10 & 0 & 0 & 12.5 \\ \hline
& & & & & & & & & Avg \# \\
& & Dose 1 & Dose 2 & Dose 3 & {\bf Dose 4} & Dose 5 & Dose 6 & Overtoxic & DLTs\\ \hline
{\bf Scenario 2} & P(DLT) & 0.19 & {\bf 0.30} & 0.44 & 0.59 & 0.72 & 0.83 & -- & -- \\
& CRM & 0.20 & {\bf 0.61} & 0.15 & 0 & 0 & 0 & 0.04 & 13.0 \\
& rCRM 1 & 0.20 & {\bf 0.62} & 0.14 & 0 & 0 & 0 & 0.04 & 13.2 \\
& rCRM 2 & 0.20 & {\bf 0.64} & 0.13 & 0 & 0 & 0 & 0.03 & 13.3 \\ \hline
& & & & & & & & & Avg \# \\
& & Dose 1 & Dose 2 & Dose 3 & {\bf Dose 4} & Dose 5 & Dose 6 & Overtoxic & DLTs\\ \hline
{\bf Scenario 3} & P(DLT) & 0.01 & 0.03 & 0.05 & 0.10 & 0.18 & {\bf 0.30} & -- & -- \\
& CRM & 0 & 0 & 0 & 0.01 & 0.32 & {\bf 0.67} & 0 & 8.6 \\
& rCRM 1 & 0 & 0 & 0 & 0.01 & 0.30 & {\bf 0.69} & 0 & 8.5 \\
& rCRM 2 & 0 & 0 & 0 & 0 & 0.27 & {\bf 0.72} & 0 & 8.4 \\ \hline
& & & & & & & & & Avg \# \\
& & Dose 1 & Dose 2 & Dose 3 & {\bf Dose 4} & Dose 5 & Dose 6 & Overtoxic & DLTs\\ \hline
{\bf Scenario 4} & P(DLT) & 0.04 & 0.11 & {\bf 0.30} & 0.59 & 0.83 & 0.94 & -- & -- \\
& CRM & 0 & 0.14 & {\bf 0.80} & 0.06 & 0 & 0 & 0 & 13.1 \\
& rCRM 1 & 0 & 0.14 & {\bf 0.80} & 0.06 & 0 & 0 & 0 & 13.5 \\
& rCRM 2 & 0 & 0.13 & {\bf 0.80} & 0.06 & 0 & 0 & 0 & 13.5 \\ \hline
& & & & & & & & & Avg \# \\
& & Dose 1 & Dose 2 & Dose 3 & {\bf Dose 4} & Dose 5 & Dose 6 & Overtoxic & DLTs\\ \hline
{\bf Scenario 5} & P(DLT) & 0.02 & 0.04 & 0.08 & 0.16 & {\bf 0.30} & 0.49 & -- & -- \\
& CRM & 0 & 0 & 0 & 0.15 & {\bf 0.76} & 0.09 & 0 & 10.9 \\
& rCRM 1 & 0 & 0 & 0 & 0.16 & {\bf 0.74} & 0.10 & 0 & 10.9 \\
& rCRM 2 & 0 & 0 & 0 & 0.15 & {\bf 0.75} & 0.10 & 0 & 10.7 \\ \hline
& & & & & & & & & Avg \# \\
& & Dose 1 & Dose 2 & Dose 3 & {\bf Dose 4} & Dose 5 & Dose 6 & Overtoxic & DLTs\\ \hline
{\bf Scenario 6} & P(DLT) & 0.08 & 0.14 & 0.21 & {\bf 0.30} & 0.41 & 0.54 & -- & -- \\
& CRM & 0 & 0.04 & 0.27 & {\bf 0.51} & 0.18 & 0.01 & 0 & 11.5 \\
& rCRM 1 & 0 & 0.03 & 0.26 & {\bf 0.53} & 0.18 & 0.01 & 0 & 11.5 \\
& rCRM 2 & 0 & 0.03 & 0.24 & {\bf 0.54} & 0.18 & 0 & 0 & 11.5 \\ \hline
\end{tabular}
\end{center}
\end{table}

\begin{figure}[t]
\begin{center}
\includegraphics[width = 6.5in]{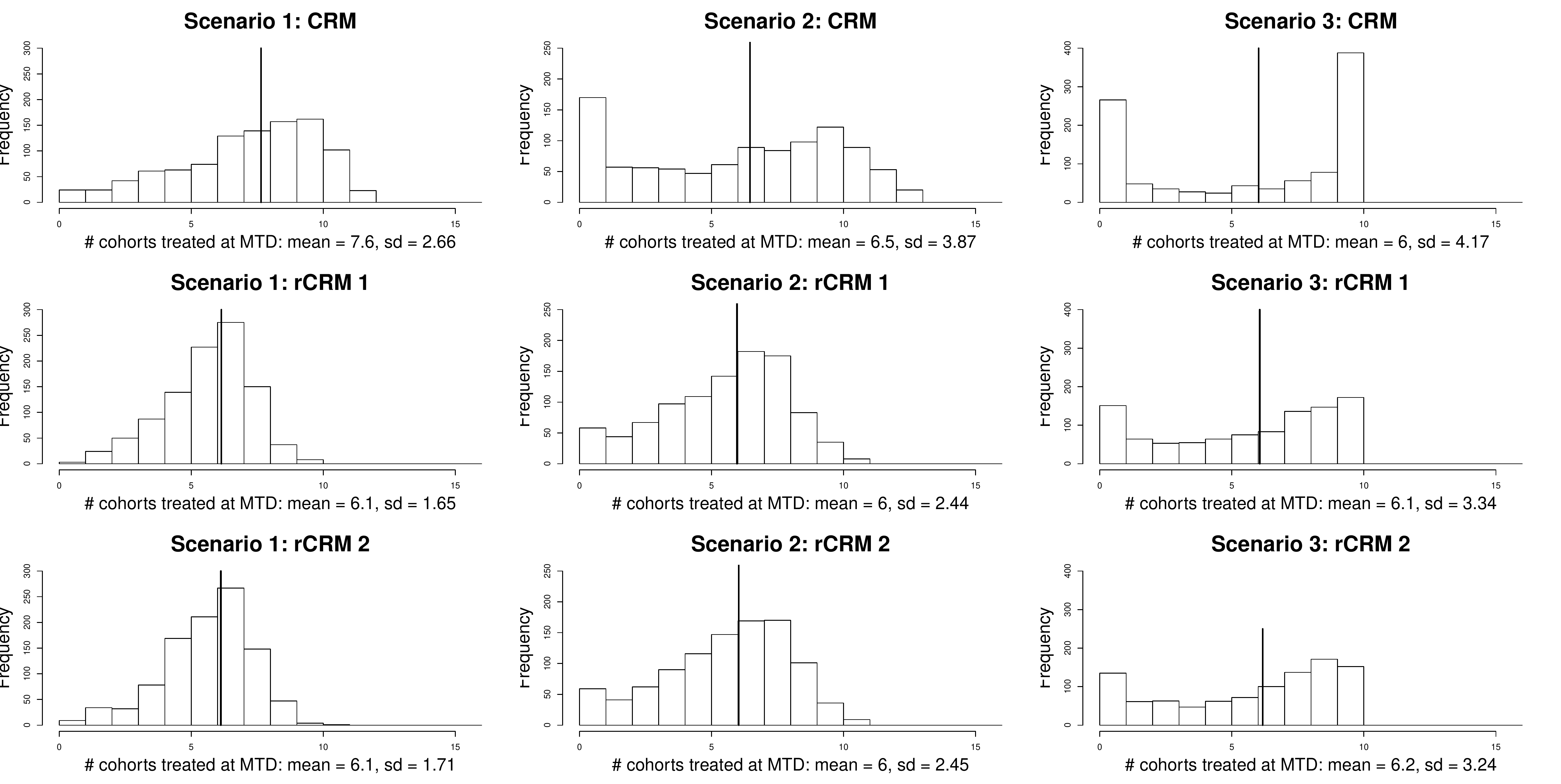}
\caption{Histogram of the number of cohorts treated at the MTD for Scenarios 1 - 3 using the CRM, rCRM 1 and rCRM 2.}
\label{fig1}
\end{center}
\end{figure}

\begin{figure}[t]
\begin{center}
\includegraphics[width = 6.5in]{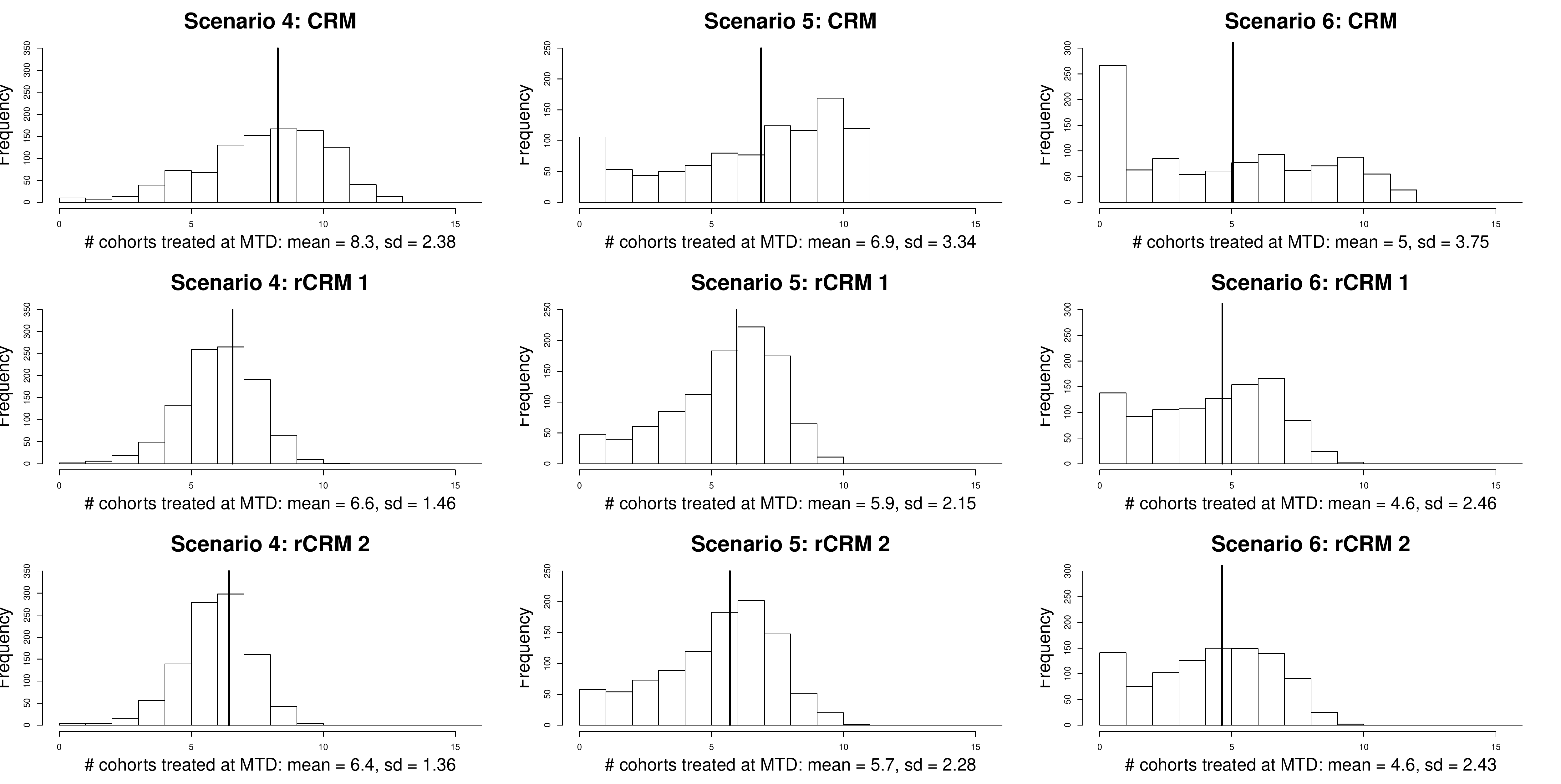}
\caption{Histogram of the number of cohorts treated at the MTD for Scenarios 4 - 6 using the CRM, rCRM 1 and rCRM 2.}
\label{fig2}
\end{center}
\end{figure}

We next consider the number of cohorts treated at the true MTD. Figures~\ref{fig1} and ~\ref{fig2} present histograms of the number of cohorts treated at the true MTD for all scenarios using the CRM, rCRM1 and rCRM2. Also provided are the mean and standard deviation of the number of cohorts treated at the true MTD for each histogram. We observe a similar pattern across the six scenarios. The rCRM1 and rCRM2 have a lower average number of cohorts treated at the true MTD than the CRM but also have a lower standard deviation than the CRM and we observe very little difference between the rCRM1 and rCRM2. Scenarios 2, 3, 5 and 6 illustrate the primary strengths of the randomized CRM. In all four cases, the standard CRM has a substantial proportion of simulated trials where 1 or 0 cohorts are treated at the true MTD. These probabilities are dramatically reduced for both the rCMR1 and rCRM2 in all four cases. In contrast, Scenarios 1 and 4 illustrate a potential weakness of the randomized CRM. In both cases, the standard CRM performs well and the rCRM1 and rCRM2 reduces the standard deviation of the number of cohorts treated at the MTD primarily by reducing the probability that a very high number of cohorts are treated at the MTD without a similar reduction in the probability that a low number of cohorts are treated at the MTD.

\section{Discussion}
\label{disc_sec}
We have proposed the rCRM as an approach to overcoming the long-memory property of the CRM. \citet{OandH13} illustrated that the standard CRM has a tendency to get ``stuck'' on a dose level with little escalation or de-escalation in the later stages of the trial. The rCRM overcomes this problem by introducing random escalation or de-escalation in cases where the standard CRM would assign a cohort the same dose as the previous cohort. We investigated two algorithms for random escalation or de-escalation. The rCRM1 randomly assigns a cohort to either escalate one dose level, de-escalate one dose level or stay the same relative to the previous cohort with probabilities proportional to the posterior probability that each dose is the MTD. In contrast, the rCRM2 randomly assigns a cohort to any dose from the minimum dose level to a dose level representing a one dose escalation relative to the maximum dose-level tried in the previous cohorts with probabilities proportional to the posterior probability that each dose is the MTD. The rCRM1 and rCRM2 correctly identified the MTD at similar rates to the standard CRM and reduced the standard deviation of the number cohorts treated at the MTD in all scenarios. Although, this did come at the cost of reducing the average number of cohorts treated at the MTD and this decrease was substantial in two of the six scenarios.

An additional advantage of the rCRM is that it allows us to incorporate the uncertainty in our estimate of the MTD into dose-finding. The standard CRM treats each cohort at the current estimate of the MTD regardless of the uncertainty in that estimate. In contrast, the rCRM considers the entire posterior distribution of the MTD and incorporates the uncertainty in our estimate into dose-finding. As a result, the rCRM is less likely to get ``stuck'' on a dose-level than the standard CRM and also has the added advantage of more thoroughly exploring the dose-response relationship of toxicity without sacrificing the probability of correctly identifying the MTD. 

We investigated the performance of two randomization schemes but other randomization schemes could also be considered. The rCRM2 randomization scheme is similar to adaptive randomization schemes that attempt to maximize the probability of treating patients with the optimal treatment from a set of multiple treatments in a randomized clinical trial \citep{BCLandM11}. Other adaptive randomization schemes have been proposed and could also be implemented in the rCRM. Alternately, a randomization scheme could be developed that incorporates a loss-function that penalizes over-dosing to develop an rCRM design with operating characteristics similar to EWOC \citep{BRandZ98}. 

Finally, we have considered random escalation and de-escalation in standard phase I clinical trials to evaluate the toxicity of a single agent. There exists a vast literature of Bayesian adaptive trial designs that consider extensions of phase I dose-escalation studies to the case of multiple agents \citep{YandY09, TMMandL03} and designs that consider efficacy, as well as toxicity, in phase I \citep{Braun02, TandC04}. These designs are also likely to have a long-memory and incorporating random escalation or de-escalation into their dose-finding algorithms could be used to overcome the long-memory property in their settings, as well.
\section*{Acknowledgements}
This work was partially supported by a research grant from Medtronic Inc.
\newpage

\bibliographystyle{biom}
\bibliography{koopmeiners_ref}

\begin{thebibliography}{}

\bibitem[\protect\citeauthoryear{Babb, Rogatko, Rogatko, and Zacks}{Babb
  et~al.}{1998}]{BRandZ98}
Babb, J., Rogatko, A., Rogatko, A., and Zacks, S. (1998).
\newblock Cancer phase {I} clinical trials: {E}fficient dose escalation with
  overdose control.
\newblock {\em Statistics in Medicine} {\bf 17,} 1103--1120.

\bibitem[\protect\citeauthoryear{Berry, Carlin, Lee, and M{\"u}ller}{Berry
  et~al.}{2011}]{BCLandM11}
Berry, S.~M., Carlin, B.~P., Lee, J.~J., and M{\"u}ller, P. (2011).
\newblock {\em Bayesian Adaptive Methods for Clinical Trials}.
\newblock Chapman \& Hall Ltd.

\bibitem[\protect\citeauthoryear{Braun}{Braun}{2002}]{Braun02}
Braun, T.~M. (2002).
\newblock The bivariate continual reassessment method: extending the {CRM} to
  phase {I} trials of two competing outcomes.
\newblock {\em Controlled Clinical Trials} {\bf 23,} 240 -- 256.

\bibitem[\protect\citeauthoryear{Gasparini and Eisele}{Gasparini and
  Eisele}{2000}]{GandE00}
Gasparini, M. and Eisele, J. (2000).
\newblock A curve-free method for phase i clinical trials.
\newblock {\em Biometrics} {\bf 56,} 609--615.

\bibitem[\protect\citeauthoryear{Goodman, Zahurak, and Piantadosi}{Goodman
  et~al.}{1995}]{GZandP95}
Goodman, S.~N., Zahurak, M.~L., and Piantadosi, S. (1995).
\newblock Some practical improvements in the continual reassessment method for
  phase {I} studies.
\newblock {\em Statistics in Medicine} {\bf 14,} 1149--1161.

\bibitem[\protect\citeauthoryear{Ivanova, Montazer-Haghighi, Mohanty, and
  D.~Durham}{Ivanova et~al.}{2003}]{IMMandD03}
Ivanova, A., Montazer-Haghighi, A., Mohanty, S.~G., and D.~Durham, S. (2003).
\newblock Improved up-and-down designs for phase i trials.
\newblock {\em Statistics in Medicine} {\bf 22,} 69--82.

\bibitem[\protect\citeauthoryear{O'Quigley, Pepe, and Fisher}{O'Quigley
  et~al.}{1990}]{OPandF90}
O'Quigley, J., Pepe, M., and Fisher, L. (1990).
\newblock Continual reassessment method: A practical design for phase 1
  clinical trials in cancer.
\newblock {\em Biometrics} {\bf 46,} pp. 33--48.

\bibitem[\protect\citeauthoryear{Oron and Hoff}{Oron and Hoff}{2009}]{OandH09}
Oron, A.~P. and Hoff, P.~D. (2009).
\newblock The k-in-a-row up-and-down design, revisited.
\newblock {\em Statistics in Medicine} {\bf 28,} 1805--1820.

\bibitem[\protect\citeauthoryear{Oron and Hoff}{Oron and Hoff}{2013}]{OandH13}
Oron, A.~P. and Hoff, P.~D. (2013).
\newblock Small-sample behavior of novel phase i cancer trial designs.
\newblock {\em Clinical Trials} {\bf 10,} 63--80.

\bibitem[\protect\citeauthoryear{Plummer}{Plummer}{2013}]{rjags}
Plummer, M. (2013).
\newblock {\em rjags: Bayesian graphical models using MCMC}.
\newblock R package version 3-10.

\bibitem[\protect\citeauthoryear{{R Core Team}}{{R Core Team}}{2013}]{Rref}
{R Core Team} (2013).
\newblock {\em R: A Language and Environment for Statistical Computing}.
\newblock R Foundation for Statistical Computing, Vienna, Austria.

\bibitem[\protect\citeauthoryear{Storer}{Storer}{1989}]{Storer89}
Storer, B.~E. (1989).
\newblock Design and analysis of phase {I} clinical trials.
\newblock {\em Biometrics} {\bf 45,} pp. 925--937.

\bibitem[\protect\citeauthoryear{Stylianou and Flournoy}{Stylianou and
  Flournoy}{2002}]{SandF02}
Stylianou, M. and Flournoy, N. (2002).
\newblock Dose finding using the biased coin up-and-down design and isotonic
  regression.
\newblock {\em Biometrics} {\bf 58,} 171--177.

\bibitem[\protect\citeauthoryear{Thall and Cook}{Thall and
  Cook}{2004}]{TandC04}
Thall, P.~F. and Cook, J.~D. (2004).
\newblock Dose-finding based on efficacy/toxicity trade-offs.
\newblock {\em Biometrics} {\bf 60,} 684--693.

\bibitem[\protect\citeauthoryear{Thall, Millikan, Mueller, and Lee}{Thall
  et~al.}{2003}]{TMMandL03}
Thall, P.~F., Millikan, R.~E., Mueller, P., and Lee, S.-J. (2003).
\newblock Dose-finding with two agents in phase i oncology trials.
\newblock {\em Biometrics} {\bf 59,} 487--496.

\bibitem[\protect\citeauthoryear{Yin and Yuan}{Yin and Yuan}{2009}]{YandY09}
Yin, G. and Yuan, Y. (2009).
\newblock A latent contingency table approach to dose finding for combinations
  of two agents.
\newblock {\em Biometrics} {\bf 65,} 866--875.

\end{thebibliography}
\end{document}